\documentclass[fleqn,usenatbib]{mnras}

\usepackage{newtxtext,newtxmath}
\usepackage{caption}
\usepackage{subcaption}

\usepackage[T1]{fontenc}

\DeclareRobustCommand{\VAN}[3]{#2}
\let\VANthebibliography\thebibliography
\def\thebibliography{\DeclareRobustCommand{\VAN}[3]{##3}\VANthebibliography}

\usepackage{graphicx}	
\usepackage{amsmath}	
\usepackage{bm} 
\usepackage{ulem}
\usepackage{balance}
\usepackage{enumitem}
\usepackage{multirow}
\usepackage[usenames,dvipsnames]{xcolor}
\usepackage{pifont}
\usepackage{orcidlink}
\usepackage{makecell}

\def\be{\begin{equation}}
\def\ee{\end{equation}}
\def\bea{\begin{eqnarray}}
\def\eea{\end{eqnarray}}

\def\S{\mathbf{S}}

\newcommand{\dvec}{\ensuremath{\boldsymbol{d}}}
\newcommand{\mvec}{\ensuremath{\boldsymbol{m}}}
\newcommand{\evec}{\ensuremath{\boldsymbol{e}}}
\newcommand{\avec}{\ensuremath{\boldsymbol{a}_{\text{fg}}}}
\newcommand{\aveci}[1]{\ensuremath{\boldsymbol{a}_{\text{fg},#1}}}
\newcommand{\nvec}{\ensuremath{\boldsymbol{n}}}
\newcommand{\svec}{\ensuremath{\boldsymbol{s}}}
\newcommand{\fvec}{\ensuremath{\boldsymbol{f}}}
\newcommand{\Smat}{\ensuremath{\mathbf{S}}}
\newcommand{\Fmat}{\ensuremath{\mathbf{F}}}
\newcommand{\Gmat}{\ensuremath{\mathbf{G}}}
\newcommand{\Emat}{\ensuremath{\mathbf{E}}}
\newcommand{\Nmat}{\ensuremath{\mathbf{N}}}
\newcommand{\Einv}{\ensuremath{\mathbf{E}^{-1}}}
\newcommand{\Ninv}{\ensuremath{\mathbf{N}^{-1}}}

\newcommand{\tmat}{\ensuremath{\mathbf{T}}}

\newcommand{\hydrapspec}{\texttt{hydra-pspec}}

\newcommand{\hpx}{\texttt{HEALPix}}

\newcommand{\figref}[1]{Figure {#1}}
\renewcommand{\eqref}[1]{Equation {#1}}
\newcommand{\secref}[1]{Section {#1}}

\usepackage{geometry}
\hypersetup{
    colorlinks = true,
    citecolor = {MidnightBlue},
    linkcolor = {BrickRed},
    urlcolor = {BrickRed}
}

\defcitealias{kennedy23}{K23}

\author[J.~Burba et al.]{
Jacob Burba$^{1}$\thanks{E-mail: jacob.burba@manchester.ac.uk}\,\orcidlink{0000-0002-8465-9341},~
Philip Bull$^{1,2}$\,\orcidlink{0000-0001-5668-3101},~
Michael J.\ Wilensky$^{1}$\,\orcidlink{0000-0001-7716-9312},~
Fraser Kennedy$^{3}$\,\orcidlink{0000-0002-5883-6543},~
Hugh Garsden$^{1}$\,\orcidlink{0009-0001-3949-9342},
\newauthor
Katrine A.\ Glasscock$^{1}$\,\orcidlink{0000-0001-6894-0902},~
\\
$^{1}$Jodrell Bank Centre for Astrophysics, University of Manchester, Manchester, M13 9PL, United Kingdom \\
$^{2}$Department of Physics and Astronomy, University of Western Cape, Cape Town 7535, South Africa \\
$^{3}$Astronomy Unit, Queen Mary University of London, Mile End Road, London E1 4NS, United Kingdom
}
\date{Accepted XXX. Received YYY; in original form ZZZ}

\pubyear{2023}

\begin{document}
\label{firstpage}
\pagerange{\pageref{firstpage}--\pageref{lastpage}}

\title[Sensitivity of Bayesian 21~cm power spectrum estimation to foreground model errors]{Sensitivity of Bayesian 21~cm power spectrum estimation to foreground model errors}
\maketitle

\begin{abstract} 
Power spectrum estimators are an important tool in efforts to detect the 21~cm brightness temperature fluctuations from neutral hydrogen at early times. An initial detection will likely be statistical in nature, meaning that it will not be possible to make a coherent map of the brightness temperature fluctuations; instead, only their variance will be measured against a background of noise and residual systematic effects. Optimal Quadratic Estimator (OQE)-based methods often apply an inverse covariance weighting to the data.  However, inaccurate covariance modelling can lead to reduced sensitivity and, in some cases, severe signal loss. We recently proposed a Bayesian method to jointly estimate the 21~cm fluctuations, their power spectrum, and foreground emission. Instead of requiring a fixed a priori estimate of the covariance, we estimate the covariance as part of the inference. Choices of parametrization, particularly of the foregrounds, are subject to model errors and could lead to biases and other ill effects if not properly controlled. In this paper, we investigate the effects of inaccurate foreground models on 21~cm power spectrum recovery.  Using simulated visibilities, we find that, even in the most extreme scenarios tested, our approach is capable of recovering 21~cm delay power spectrum estimates consistent with a known input signal for delays $\gtrsim300$ ns ($\sim$88\% of the available Fourier modes).  This is true even when using foreground models derived from modified foreground catalogs containing spatial and spectral perturbations at the quoted level of uncertainty on our foreground catalogs.
\end{abstract}




\section{Introduction}
\label{sec:intro}

The low-frequency radio sky is dominated by synchrotron and free-free emission from our own Galaxy. While observations on small angular scales, e.g. of distant galaxies, are relatively untroubled by it, attempts to measure extragalactic emission on large angular scales is severely impacted. This includes attempts to measure the radio background, including a potential excesses suggested by the ARCADE2 \citep{fixsen11}, LWA \citep{dowell18}, and EDGES \citep{bowman18} experiments for example, and the cosmological fluctuations in the 21~cm brightness temperature field at both low and high redshifts \citep{paciga13, mertens20, trott20, heraps23, cunnington23}. In the latter case, the foreground contamination is particularly troublesome due to the high dynamic range between the synchrotron emission and the 21~cm fluctuations, which can be as large as four or five orders of magnitude in temperature (see e.g.\ \citet{mertens20, trott20, heraps23} and sources within). While Galactic synchrotron is expected to have a mostly smooth, power-law-like frequency spectrum \citep{cmb-fgs, liu-tegmark, gmoss, gsm, gsmref, synch-fgs, planck16, planck18-fgs}, in contrast to the complex frequency structure of the cosmic 21~cm field, modulation of the foreground signal by calibration errors and the complicated, frequency-dependent response of typical radio telescopes results in a spurious spectrally-varying component of the measured data that can be hard to disentangle (see e.g.\ \citet{morales-wedge, Barry:2016cpg, byrne19, Orosz:2018avj}). Because of the high dynamic range, this spurious component can easily bury the true 21~cm unless extremely precise calibration and instrumental modelling can be applied.

Numerous methods of handling foreground contamination have been proposed in the context of 21~cm experiments, ranging from methods that segregate the data into contaminated and uncontaminated regions in Fourier space (so-called `foreground avoidance'), to `blind' methods that use the spectral smoothness and high variance of the foreground signal to construct a filter, to `semi-blind' forward-modelling approaches that aim to reconstruct a high-fidelity model of the sky (see \citet{liu-shaw} and sources within for a comprehensive review of current FG mitigation strategies).

Each method has its strengths and weaknesses, but there are some common limitations. One is the potential for signal loss, where uncertain or overly flexible foreground model components or filter modes erroneously absorb some of the 21~cm signal as well, particularly on large radial scales (low $k_\parallel$). This biases the recovered signal, necessitating the application of a transfer function to correct (de-bias) the measured power spectrum of the data. Such transfer functions are usually constructed by passing known simulated signals through the same foreground filtering steps to examine where the signal loss has occurred \citep{h1c-validation, heraps22, heraps23}.

Another issue is handling the additional complexity of the foreground component due to calibration errors and instrumental artifacts. These are difficult to model precisely, and can introduce further undesirable complications such as coupling between different Fourier modes. As one example, Principal Component Analysis (PCA) filters only typically need to subtract the first few PCA modes when applied to simulated autocorrelation (single-dish) 21~cm data, whereas real data require more like 30 modes to be subtracted \citep{Spinelli:2021emp, 2023MNRAS.518.6262C}. Similarly, interferometers like HERA, pursuing a foreground avoidance strategy, need to add an additional buffer region around the contaminated `foreground wedge' region in Fourier space to avoid foreground power that leaks out due to instrumental artifacts such as non-redundant primary beams \citep{Orosz:2018avj, kim2023impact}.

Finally, real data tend to have substantial masking/flagging due to radio-frequency interference (RFI) and other data quality issues. This can be challenging for many foreground removal methods, as well as for power spectrum estimation -- gaps in the data cause ringing and mode-coupling artifacts when the data are Fourier transformed for example \citep{offringa2019, mertens20, trott20, heraps23, wilensky23}. Inhomogeneous noise, caused by a greater amount of flagging in some parts of the data than others, can also have adverse effects \citep{offringa2019, 2022MNRAS.510.5023W}. Approaches to mitigating these issues can cause their own complications; for instance, in-painting flagged data with a model can introduce biases, while attempts to take into account the structure of the noise are subject to various pitfalls of empirical estimates of covariance matrices from data \citep[][Feng et al. {\it in prep}]{Cheng:2018osq}.

\citet{kennedy23} (hereafter \citetalias{kennedy23}) presented a Bayesian approach for combined foreground modelling, in-painting of missing data, and power spectrum estimation. This used the combined methods of Gaussian constrained realizations (GCRs) to draw samples from very high dimensional conditional distributions for the 21~cm signal and foreground modes given (gappy/incomplete) data, and Gibbs sampling to iteratively sample from the joint posterior distribution of the 21~cm signal, foreground mode amplitudes, and the delay power spectrum of the 21~cm signal itself. The foreground modes could be constructed from any chosen set of basis functions in frequency, which were in this instance derived from a PCA decomposition of simulated foreground visibility data. \citetalias{kennedy23} showed that, for simple visibility simulations with 21~cm signal, foregrounds, and thermal noise, this method permitted the 21~cm delay spectrum to be recovered in an unbiased way, at the expense of increased variance at low delays (small $k_\parallel$) where the uncertain foreground model prevented more accurate 21~cm signal recovery.

In this paper, we address the main limitations of the analysis in \citetalias{kennedy23}, with the aim of validating this method -- called \hydrapspec\ -- for use on real data, with all the complications that that entails (see \secref{\ref{sec:math}} for a high level summary of our approach).  First, we use substantially more realistic simulations of visibility data, including an extensive point source catalog, a diffuse sky model, and a realistic primary beam model (\secref{\ref{sec:sims}}). Next, we address imperfections in the construction of the foreground model basis functions due to uncertainties in the foreground frequency-frequency covariance matrix that was used in the PCA decomposition (\secref{\ref{sec:fg}}). We perform a variety of tests of foreground model imperfections, including the effect of missing/unresolved sources, and spatially- and spectrally-dependent deviations from the assumed diffuse model frequency spectrum (\secref{\ref{sec:results}}). Finally, we touch on the interplay between the primary beam of the interferometer and temporal correlations in the signal and foregrounds, which are not currently modelled by our method (\secref{\ref{sec:time-correlations}}).  We conclude and present future directions in \secref{\ref{sec:conclusions}}.

\section{Bayesian Power Spectrum Estimation}
\label{sec:math}

In this section, we present a high-level overview of our Bayesian approach to 21~cm delay power spectrum estimation, \hydrapspec. This framework was first presented in \citetalias{kennedy23}, where simplified foreground and signal models were used to demonstrate the properties of the method.  For more detail on the mathematics behind our approach, we refer the reader to \citetalias{kennedy23}.  Notation-wise, unless otherwise specified, we will use lowercase bold symbols (e.g.\ \svec) to refer to vectors and bold capital letters to refer to matrices (\Smat).  We have made two small changes to the notation defined in \citetalias{kennedy23} and have added footnotes describing these changes where relevant.

We begin by assuming that our data (\dvec) is a sum of EoR (\evec), FGs (\fvec), and Gaussian distributed noise (\nvec), i.e.
\begin{equation}
\dvec = \evec + \fvec + \nvec
\end{equation}
Specifically, the data are complex-valued visibilities from an interferometer per frequency, time, and baseline
\begin{equation}
    V_{mn}(\nu, t) = w(\nu, t)\left[ e_{mn}(\nu, t) + f_{mn}(\nu, t) + n_{mn}(\nu, t) \right]
\end{equation}
where indices $m, n$ label the antennas forming a given baseline, $\nu$ and $t$ label frequency and observing time (local sidereal time, LST), respectively, and $\boldsymbol{w}$ is a mask vector of 1s and 0s for flagged and unflagged data, respectively.  Assuming each of these components to be independent and mean zero, the data covariance can be written as
\begin{align}
    \mathbf{C} &= \left< \dvec\dvec^\dagger \right> = \left< \evec\evec^\dagger \right> + \left< \fvec\fvec^\dagger \right> + \left< \nvec\nvec^\dagger \right> = \Emat + \Fmat + \Nmat\,,
\end{align}
where angle brackets denote an ensemble average and \Emat, \Fmat, and \Nmat\ are the covariance matrices of the EoR, FGs, and noise, respectively.

In this work, we follow ``Scheme 2'' from \citetalias{kennedy23} where we construct the foregrounds in our model via a set of spectral templates (\Gmat) and coefficients (\avec) such that
\begin{equation}
    \fvec = \Gmat\avec
\end{equation}
In the equation above, \Gmat\footnote{In \citetalias{kennedy23}, this matrix was referred to as $\mathbf{g}_j$.  This notation was adopted from \citet{eriksen2008} but we have changed it here to make it clear, given our choice of notation, that \Gmat\ is a matrix whereas $\mathbf{g}_j$ could be misinterpreted as a vector due to its lowercase symbol.  For posterity, we also wish to point out that \Gmat\ was previously used in \citetalias{kennedy23} for the Weiner filter.  To be pedantic, here we use \Gmat\ to represent a column matrix comprised of spectral templates for the FG model.} is a matrix whose columns are comprised of a subset of leading principle components obtained from the FG covariance matrix \Fmat\ via eigendecomposition (see \secref{\ref{sec:fg-model}} for more detail) and \avec\ is a vector of corresponding coefficients.  This technique was motivated by the work done in \citet{eriksen2008} which used a similar approach for component separation in analyzing cosmic microwave background data.  Our data model in this scheme takes the form
\begin{equation}
    \mvec = \evec + \Gmat\avec
    \label{eq:model-vector}
\end{equation}

We wish to draw samples from the joint posterior $p(\evec, \Emat, \avec | \Gmat, \Nmat, \dvec)$.  In \hydrapspec, we use the technique of Gibbs sampling \citep{gibbs} to obtain samples from this posterior by iteratively drawing samples from a set of conditional distributions\footnote{This assumes that the joint probability density is strictly positive over the span of each variable.  In this case, the joint probability density is fully specified by the full set of conditional distributions.}.  In each iteration of our sampler, we have two Gibbs steps
\begin{align}
    \evec_{i+1}, \aveci{i+1} &\leftarrow p(\evec_i, \aveci{i}| \Emat_i, \Gmat, \Nmat, \dvec)
    \label{eq:signal-gibbs-1}\\
    \Emat_{i+1} &\leftarrow p(\Emat_i | \evec_{i+1})
    \label{eq:signal-gibbs-2}
\end{align}
where $\leftarrow$ in the above representation implies sampling.  In words, we first draw an updated sample of the EoR and FG models conditioned on the EoR covariance from the previous iteration and then draw an updated sample of the EoR covariance conditioned on the updated EoR and FG models.

For the first Gibbs step (\eqref{\ref{eq:signal-gibbs-1}}), we consider a posterior conditioned on known covariance information of the EoR (\Emat) and noise (\Nmat) of the form
\begin{align}
    p(\evec, \avec& | \Emat, \Gmat, \Nmat, \dvec) \propto p(\dvec | \evec, \avec, \Gmat, \Nmat) p(\evec | \Emat)
    \nonumber\\
    \propto
        &\exp\left[ - (\dvec - (\evec + \Gmat\avec))^\dagger\Ninv(\dvec - (\evec + \Gmat\avec)) \right]
        \nonumber\\
        &\times\exp\left( -\evec^\dagger\Einv\evec \right)
    \label{eq:joint-posterior}
\end{align}
In the above equation, we have assumed a flat prior on the FGs and that \evec\ is a single EoR spectrum at a given Local Sidereal Time (LST).  We assume that each time sample in our visibility data is independent (see \secref{\ref{sec:time-correlations}} for more information).  We thus draw samples of the EoR and FG signals for each LST independently but parallelize over the time axis to draw samples at different LSTs simultaneously.  To obtain samples of \evec\ and \avec at each LST, we employ the technique of Gaussian Constrained Realizations (GCRs).  The GCR approach utilizes the fact that \eqref{\ref{eq:joint-posterior}} is a multivariate Gaussian.  We can draw multiple realizations from this distribution using the combination of the Wiener filter solution (maximum a posteriori solution) plus Gaussian random variables scaled by the appropriate covariances \citep{eriksen2008}.  The Wiener filter solution can be obtained by differentiating \eqref{\ref{eq:joint-posterior}} and solving for the \evec\ and \avec\ which maximize the posterior.  This results in the linear system given by
\begin{equation}
\label{eq:wiener-filter}
    \begin{bmatrix}
        \Einv + \Ninv    & \Ninv\Gmat \\[4pt]
        \Gmat^\dagger\Ninv & \Gmat^\dagger\Ninv\Gmat
    \end{bmatrix}
    \begin{bmatrix}
        \evec \\[4pt]
        \avec
    \end{bmatrix}
    =
    \begin{bmatrix}
        \Ninv\dvec\\[4pt]
        \Gmat^\dagger\Ninv\dvec
    \end{bmatrix}
\end{equation}
To draw samples from the posterior in \eqref{\ref{eq:joint-posterior}}, we add zero-mean, unit-variance Gaussian random vectors ($\boldsymbol{\omega}_e$, $\boldsymbol{\omega}_n)$\footnote{In \citetalias{kennedy23}, $\boldsymbol{\omega}_e$ and $\boldsymbol{\omega}_n$ were defined as $\boldsymbol{\omega}_0$ and $\boldsymbol{\omega}_1$, respectively.  These subscripts (0, 1) have been updated with symbols ($e$, $d$) to make it clear that $\boldsymbol{\omega}_e$ gets scaled by \Emat\ while $\boldsymbol{\omega}_n$ gets scaled by \Nmat.} scaled by the EoR and noise covariances to the right-hand-side of the linear system in \eqref{\ref{eq:wiener-filter}}.  This results in the following linear system
\begin{equation}
\label{eq:gcr-og}
    \begin{bmatrix}
        \Einv + \Ninv    & \Ninv\Gmat \\[4pt]
        \Gmat^\dagger\Ninv & \Gmat^\dagger\Ninv\Gmat
    \end{bmatrix}
    \begin{bmatrix}
        \evec \\[4pt]
        \avec
    \end{bmatrix}
    =
    \begin{bmatrix}
        \Ninv\dvec  + \Emat^{-1/2}\boldsymbol{\omega}_e + \Nmat^{-1/2}\boldsymbol{\omega}_n\\[4pt]
        \Gmat^\dagger\Ninv\dvec + \Gmat^\dagger\Nmat^{-1/2}\boldsymbol{\omega}_n
    \end{bmatrix}
\end{equation}
Note that in the above system of equations, there are two occurrences of $\boldsymbol{\omega}_n$.  This symbol refers to the same random draw, as opposed to two independent draws, and is merely used in two locations in the linear system.  In general, the EoR covariance matrix in frequency space, \Emat, is non-trivial and not necessarily diagonal.  To avoid numerical issues associated with inverting non-diagonal matrices, we use a modified form of Equation 22 from \citetalias{kennedy23} where we multiply the first row of the GCR system by $\Emat$, i.e.
\begin{equation}
\label{eq:gcr}
    \begin{bmatrix}
        \mathbf{1} + \Emat\Ninv    & \Emat\Ninv\Gmat \\[4pt]
        \Gmat^\dagger\Ninv & \Gmat^\dagger\Ninv\Gmat
    \end{bmatrix}
    \begin{bmatrix}
        \evec \\[4pt]
        \avec
    \end{bmatrix}
    =
    \begin{bmatrix}
        \Emat\Ninv\dvec  + \Emat^{1/2}\boldsymbol{\omega}_e + \Emat\Nmat^{-1/2}\boldsymbol{\omega}_n\\[4pt]
        \Gmat^\dagger\Ninv\dvec + \Gmat^\dagger\Nmat^{-1/2}\boldsymbol{\omega}_n
    \end{bmatrix}
\end{equation}
We assume, however, that \Nmat\ is diagonal and obtaining \Ninv\ is thus trivial.  As previously mentioned, we sample \evec\ and \avec\ for each LST independently and in parallel.  For $N_t$ LSTs in the data vector, we thus obtain a set of $N_t$ solution vectors for \evec\ and \avec\ per iteration.

In practice, most visibility data observed by an Earth-based interferometer will be corrupted with artificial radio frequency interference (RFI).  The presence of this RFI in our observations requires us to flag certain times and/or frequencies where we observe the RFI.  This results in gaps in the data.  Our GCR approach here can compensate for this flagging pattern and is capable of drawing samples within these flagged regions where we have no data.  In the case of a set of flags $\boldsymbol{w}$ (or equivalently weights), we use an amended form of the inverse noise covariance given by
\begin{equation}
    \Ninv_{\boldsymbol{w}} = \boldsymbol{w}\boldsymbol{w}^T\circ\Ninv
    \label{eq:flag-noise-cov}
\end{equation}
where $\circ$ implies element-wise multiplication.  By scaling $\Ninv$ in this way, we automatically account for zeroing the flagged regions in the data vector in \eqref{\ref{eq:gcr}}.

Note that, although the signal covariance does not typically go to zero inside the flagged regions and thus takes over in the absence of noise covariance information, we are {\it not} simply drawing samples directly from the prior inside the flagged regions. The signal realization within the flagged region must be continuous and consistent with the data surrounding the flagged region.  The generated samples within flagged regions have contributions from \Emat, \Nmat, and the data in the unflagged regions \citepalias[see][section 2.2]{kennedy23}.

\begin{figure*}
    \centering
    \includegraphics[width=0.95\linewidth]{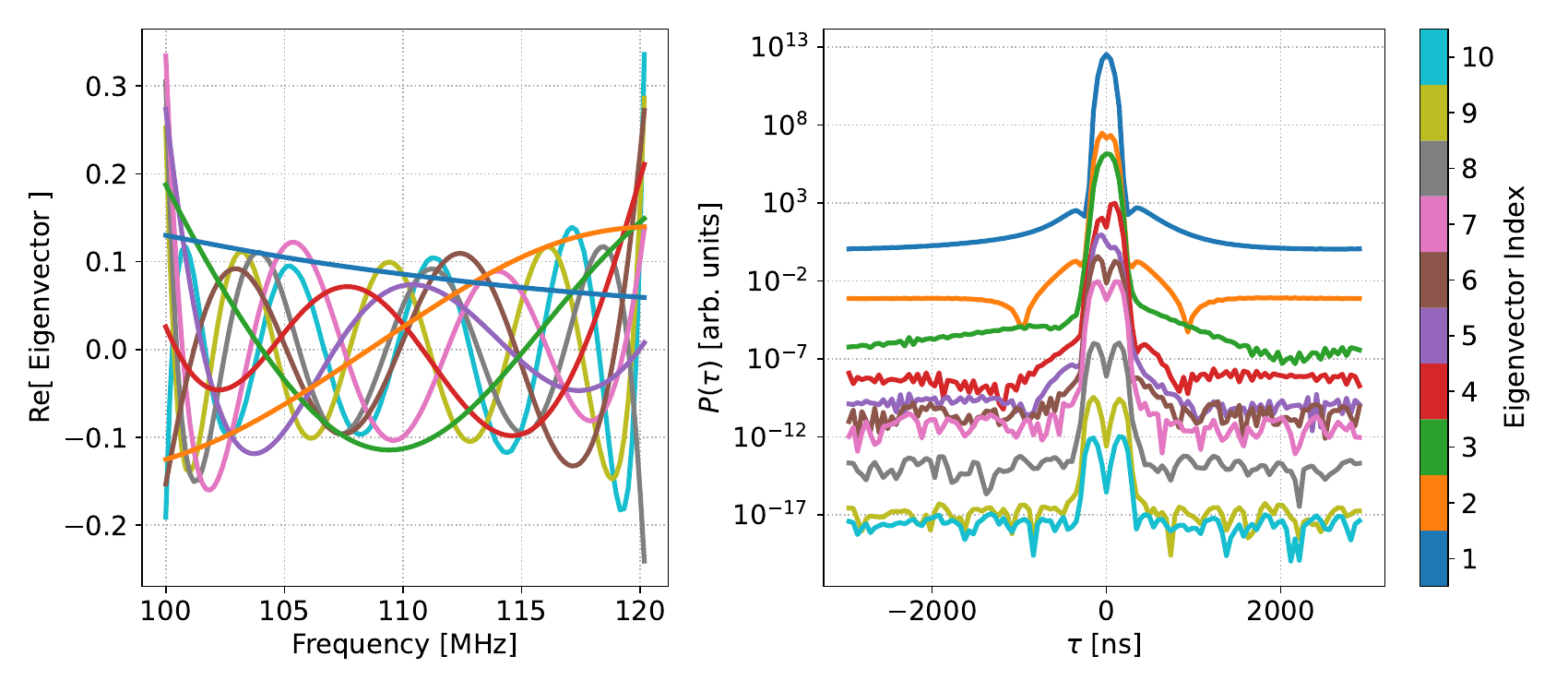}
    \caption{(Left) Real component of the first 10 principal components (eigenvectors) from the frequency-frequency covariance matrix of the ``true'' FGs (see \secref{\ref{sec:sims}} for more information).  The colorbar at right indexes the principal component index where the eigenvalue amplitude decreases with increasing index.  (Right) Delay power spectrum of each principal component multiplied by the corresponding eigenvalue.  While we use the first 12 principal components in our analyses, we only show the first 10 here to avoid visual clutter.}
    \label{fig:fg-model-pca}
\end{figure*}

For the second step in each Gibbs iteration (\eqref{\ref{eq:signal-gibbs-2}}), we consider the conditional distribution of the EoR covariance
\begin{equation}
    p(\Emat | \evec)
        = \frac{p(\evec | \Emat)p(\Emat)}{p(\evec)}
        \propto \frac{1}{\det(\Emat)}\exp\left( -\evec^\dagger\Einv\evec \right)
    \label{eq:inv-wishart}
\end{equation}
where $\det(\ldots)$ refers to the matrix determinant. Note that there is no square root of the determinant of factor of $1/2$ in the exponential because we are dealing with complex random variables and have written this expression in complex form \citep{gallager}.  Drawing samples of \Emat\ thus requires drawing samples from an inverse Wishart distribution.  When sampling \Emat, we make a simplifying assumption that the EoR signal covariance matrix in Fourier (delay) space is diagonal.  This is equivalent to assuming that the EoR signal is isotropic along the line-of-sight.
We thus define the delay space EoR covariance $\tilde{\Emat} = \tmat^\dagger\Emat\tmat$ as a diagonal matrix where \tmat\ represents a Fourier transform matrix from frequency ($\nu$) to Fourier space (delay, $\tau$) \citep{Parsons:2012qh}.  In this case, the diagonal entries of $\tilde{\Emat}$ contain the delay bandpowers, $P(\tau)$, and each delay bin is considered to be independent of all other delay bins.  The inverse Wishart distribution of \eqref{\ref{eq:inv-wishart}} then reduces to a set of independent, inverse gamma distributions, one for each delay bin.  The scale parameter for each delay bin, $\beta_\tau$, is calculated from the Fourier-transformed EoR model visibilities, $\tilde{\evec} = \tmat \evec$, via
\begin{equation}
    \beta_\tau = \sum_{t}^{N_t} \tilde{\evec}^*_{\tau, t, i}\tilde{\evec}_{\tau, t, i}
    \label{eq:beta-inv-gamma}
\end{equation}
where $\tau$ is the central delay of each delay bin, $t$ indexes the time axis (LST), and $i$ indexes the Gibbs iteration number.  The resulting shape parameter, $\alpha$, for each inverse gamma distribution is the number of times in the visibilities minus 1, i.e. $\alpha = N_t - 1$.  Under ``Scheme 2'', our Gibbs steps per iteration are thus
\begin{align}
    \evec_{i+1}, \aveci{i+1}
        &\leftarrow p(\evec_i, \aveci{i} | \Emat_i, \Gmat, \Nmat, \dvec);
        \\
    P(\tau)_{i+1}
        &\leftarrow \text{Inv-Gamma}(N_t - 1, \beta_\tau)
\end{align}
Once we obtain the delay power spectrum, we can then assign the delay bandpowers $P(\tau)_{i+1}$ to the diagonal of $\tilde{\Emat}_{i+1}$ and perform an inverse-Fourier transform to obtain our sample of $\Emat_{i+1}$.

The FG signals are assumed to be smooth and thus the bulk of their power is concentrated at low-delay modes.  We expect the EoR signal to have structure at all delays, however (see e.g.\ \citet{morales-hewitt-2004}).  There will thus be a degeneracy between the EoR and FG signals at low delays.  To control this degeneracy, we incorporate a truncated log-uniform prior on the central 7 delay bins.  The truncation is implemented using rejection sampling within a chosen interval.  This log-uniform prior acts to increment the shape parameter of the inverse gamma distribution, i.e. $\alpha' = \alpha + 1 = N_t$.  The rejection sampling only permits samples drawn from this inverse gamma distribution within the interval [0.1, 2] in delay power spectrum units.  For our purposes here, the input EoR delay power spectrum is known because we are simulating visibilities.  When analyzing real data, the EoR delay power spectrum will not be known a priori and care must be taken to choose the appropriate prior and bounds, e.g.\ the prior range can be set based on the expected delay power spectrum of the noise when the signal-to-noise-ratio (SNR) is low.  Theoretical EoR power spectra can also be used to inform the priors placed on these low delay bins.  We do not consider the scenario of an unknown EoR signal here however, and leave an investigation of the appropriate choice of prior in this case to future work.

\subsection{Constructing a Foreground Model}
\label{sec:fg-model}

For the analyses performed here, the FG models are obtained via the principal components of the frequency-space covariance matrix of the FG only visibilities.  In detail, we take an array of foreground only visibilities for each time and frequency in our simulation and compute a frequency-frequency covariance matrix, $\mathbf{F}(\nu, \nu')$.  Note that because visibilities are complex quantities, $\mathbf{F}(\nu, \nu')$ is also complex.  We then perform a principal component analysis (PCA) using the eigenvalues and eigenvectors of $\mathbf{F}(\nu, \nu')$ and use all eigenvectors with non-negligible eigenvalues (e.g. $\geq10^{-12}$).  We end up with a set of complex eigenvectors which form the basis vectors of the FG model (\Gmat\ in \eqref{\ref{eq:model-vector}}).  In all of the analyses presented here, we used the first 12 eigenvectors as the FG model basis vectors, i.e. $N_{\text{fgmodes}}=12$.  This decision was made based upon the eigenspectrum of $\mathbf{F}(\nu, \nu')$, as a plateau is reached at 12 principal components, beyond which the eigenvalues have negligible amplitudes\footnote{We have also tested the case where we use all eigenvectors in our FG model and see no difference between the recovery using only the leading 12 eigenvectors.}.  For reference, \figref{\ref{fig:fg-model-pca}} shows the real part of the first 10 components in the left hand panel and their delay power spectra (multiplied by the corresponding eigenvalue) in the right hand panel.  In all delay power spectrum plots in this work, we apply a Blackman-Harris taper prior to forming delay power spectra to enforce periodicity of the signals prior to Fourier transforming.  Additionally, because we have not yet included the full cosmological power spectrum normalization into our analysis pipeline, all power spectrum plots will be labeled with arbitrary units, or arb. units for short.  Note that this is not an issue for the purposes of our work here.  We are focused only on the relative amplitudes of the signal components in our simulated data, e.g. the SNR.

\begin{figure*}
    \centering
    \includegraphics[width=\linewidth]{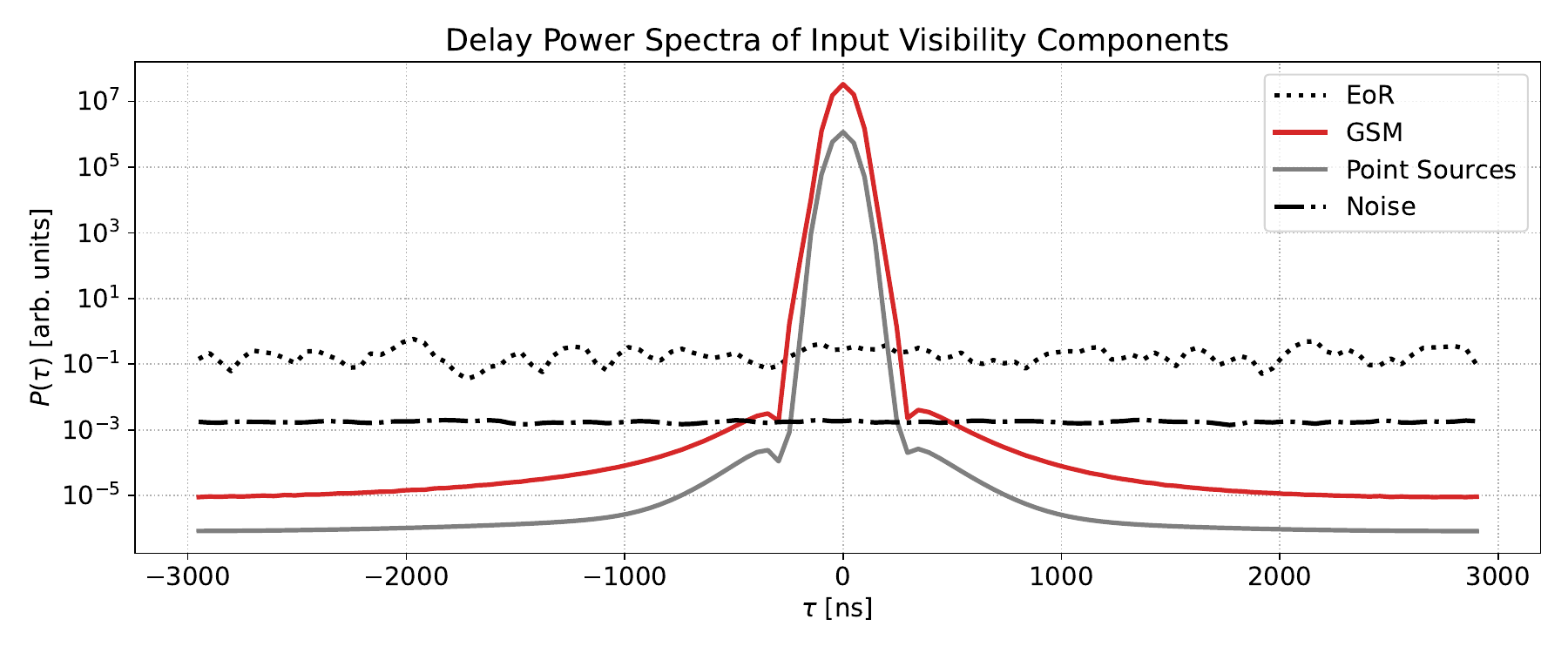}
    \caption{Delay power spectra (DPS, $y$-axis) of each component in the simulated visibilities described in \secref{\ref{sec:sims}} vs delay (Fourier dual of frequency, $x$-axis).  The mock EoR DPS is seen as the dotted black line.  The SNR was chosen to be 10 in the visibilities which corresponds to an SNR of 100 in the power spectrum.  The DPS of the noise is seen as the black dash-dot line.  The diffuse (GSM) and point source FG components are seen as the solid red and gray lines, respectively.  The amplitude of the mock EoR signal was chosen to be highly detectable at intermediate to high delay.}
    \label{fig:dps-sim-components}
\end{figure*}

\section{Simulated Data}
\label{sec:sims}

The data analyzed were comprised of visibility simulations performed with \texttt{pyuvsim}\footnote{\url{https://github.com/RadioAstronomySoftwareGroup/pyuvsim}}, a highly accurate and thoroughly tested visibility simulator \citep{pyuvsim, h1c-validation, lanman-murray-jacobs}.  For this work, we focused on the analysis of data from a HERA-like instrument and consider the case of a single 14.6~m purely East-West orientated baseline. This is the most numerous baseline type of HERA, and so is anticipated to have the highest sensitivity.  For the beam, we used an Airy beam corresponding to dish antennas with a diameter of 14.6 m.  For the frequency axis, we simulated a 20 MHz bandwidth from 100-120 MHz with a spectral resolution of $\sim$169 kHz to match the simulations in \citetalias{kennedy23}, resulting in 120 frequency channels.  For the time axis, we simulated 2.25 hours of LST with a time cadence of 40 seconds resulting in 203 time samples.

The simulated data analyzed here is comprised of a mock EoR signal and diffuse and point source FG emission.  The mock EoR signal was generated as a set of {\tt nside=16} white noise \hpx\footnote{\url{http://healpix.sourceforge.net}} \citep{healpix} maps.  We arbitrarily chose the standard deviation for the Gaussian distribution from which the EoR maps were drawn as $\sigma=91.6$ mK.  We chose a thermal noise level to allow for a high-significance detection of this mock EoR signal at all delays (discussed further below).

For the foreground models, we used the 2016 Global Sky Model \citep[GSM;][]{gsmref} obtained as a set of {\tt nside=16} \hpx\ maps from \texttt{pygdsm}\footnote{\url{https://github.com/telegraphic/pygdsm}} for the diffuse component, and the source catalog from \citet{choudhuri21} (also used in \citetalias{kennedy23}) for the point source component.  The point source catalog was generated from the Galactic and Extra-galactic All Sky MWA Survey \citep[GLEAM;][]{hurley17} but with the gaps along the Galactic plane and at northern declinations filled in with sources randomly drawn from elsewhere on the sky \citep[see][]{choudhuri21}.

The noise added to the simulated visibilities was generated as complex white noise in the visibility domain, with a standard deviation chosen such that the signal-to-noise ratio (SNR) of the visibilities is SNR$_{\text{vis}}$=10.  Because the delay power spectrum is proportional to the visibilities squared, this corresponds to an SNR in the delay power spectrum of SNR$_{\text{DPS}}$=100.  This high SNR value was chosen to make the EoR highly detectable and represents a difficult scenario where modelling errors cannot get buried by the noise.  While we have tested datasets with SNR$\leq1$ without issue, we leave a full investigation of the effect of the SNR to future work.  \figref{\ref{fig:dps-sim-components}} shows the delay power spectrum of each simulation component for reference.  The black dotted, red solid, gray solid, and black dash-dot lines show the delay power spectrum of the EoR, diffuse (GSM), point sources, and noise, respectively.

\section{Foreground models}
\label{sec:fg}

In this paper, our goal is to observe the effects of incomplete and/or inaccurate knowledge of FGs on our power spectrum estimation technique.  To do so, we kept the input data to our pipeline fixed and tested a suite of FG models inside \hydrapspec.  The fixed input data represents the ``true'' sky as seen by our simulated instrument and were simulated from the mock EoR, point source, and diffuse FG catalogs described in \secref{\ref{sec:sims}}.  To obtain incomplete and/or inaccurate FG models, we generated a set of modified FG catalogs.  For each modified FG catalog (or combination of catalogs), we simulated a set of FG-only visibilities from which we obtain the corresponding frequency-frequency covariance, which in turn gives us a FG model (see \secref{\ref{sec:fg-model}}).  The specific FG models we tested are as follows:
\begin{enumerate}[label={\arabic*.}, align=left]
    \item True FGs
    \item Point sources only (complete catalog)
    \item Bright point sources only ($S(\nu=100\text{ MHz})\geq15$ Jy)
    \item Bright point sources only with 5\% random flux errors
    \item True point sources + diffuse with 5\% random error on large scales, spectral structure on
    \begin{enumerate}[label={\alph*.}, align=left, topsep=2pt]
        \item $\Delta\nu>$ bandwidth scales ($\xi=10$)
        \item $\Delta\nu\sim$ bandwidth scales ($\xi=1$)
        \item $\Delta\nu<$ bandwidth scales ($\xi=0.1$)
        \item $\Delta\nu\ll$ bandwidth scales ($\xi=0.01$)
    \end{enumerate}
\end{enumerate}

\begin{figure*}
    \centering
    \includegraphics[width=0.48\linewidth]{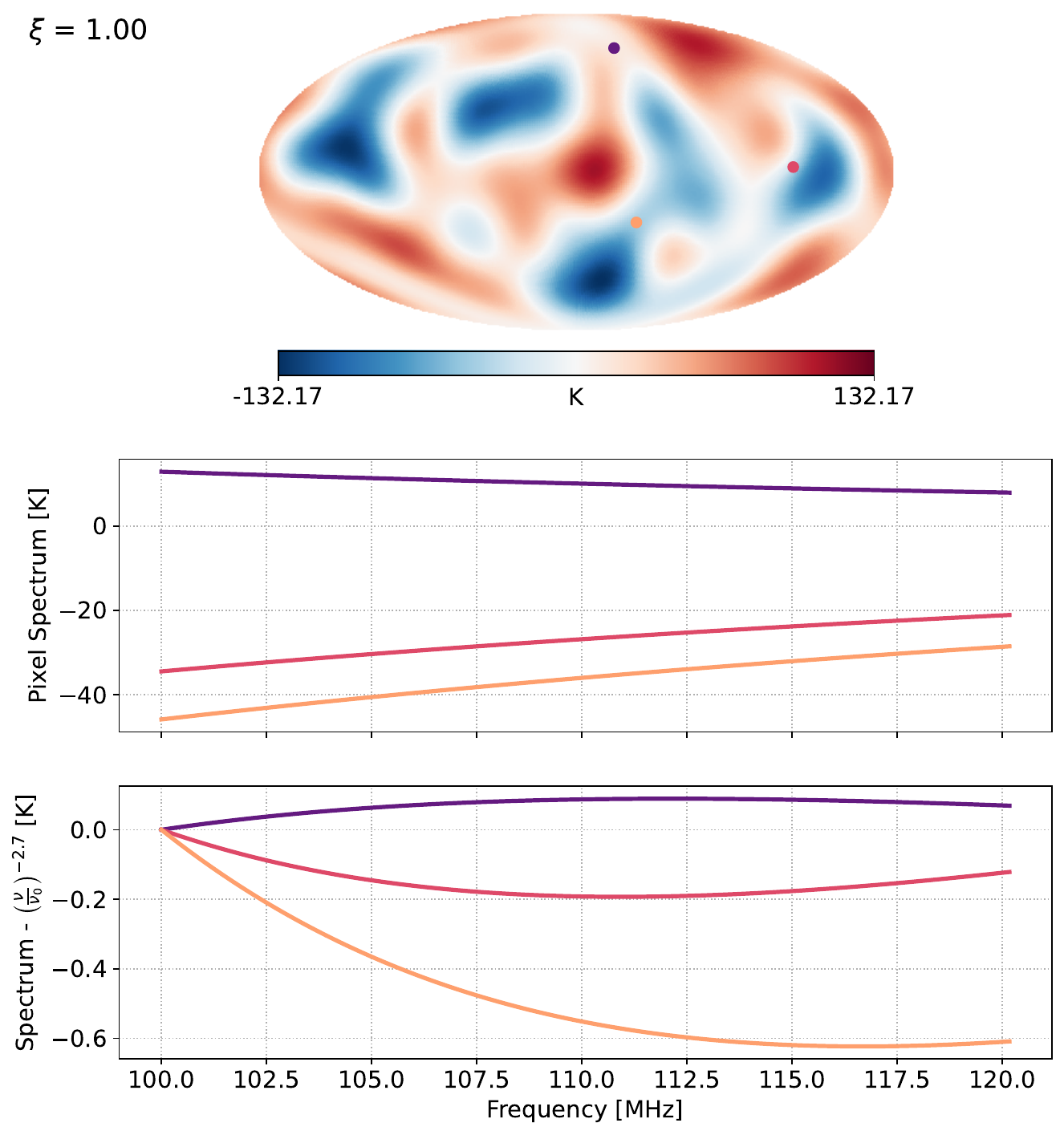}
    \includegraphics[width=0.48\linewidth]{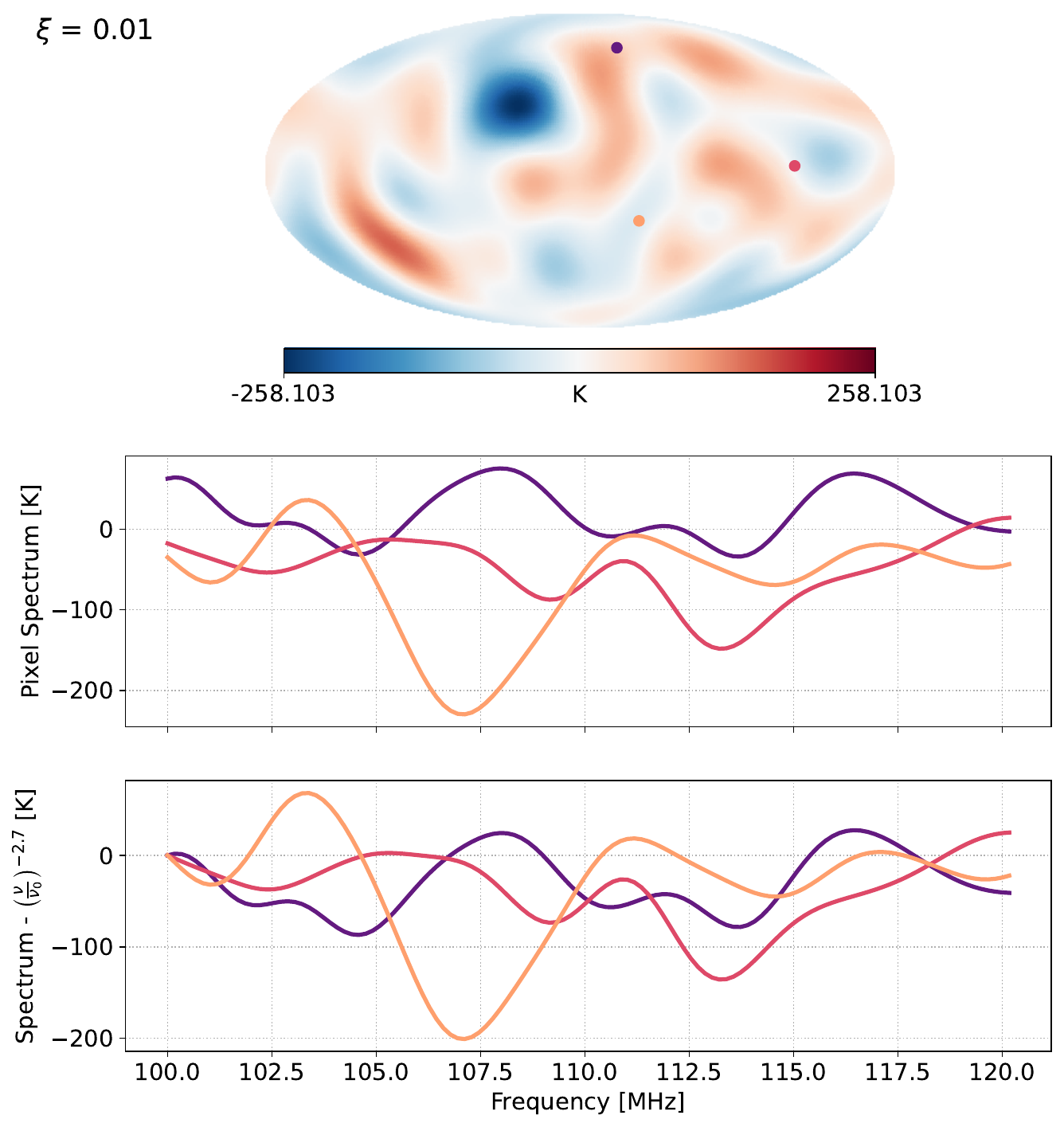}
    \caption{(Top row) \hpx\ maps of additive perturbations applied to the GSM for $\xi=1$ ($\xi_\nu\sim50$ MHz) and $\xi=0.01$ ($\xi_\nu\sim0.5$ MHz) in units of Kelvin at left and right, respectively.  Three randomly chosen pixels are highlighted in orange, red, and purple.  (Middle row) Spectrum of the three randomly chosen pixels with colors chosen to match the corresponding markers in the top plot.  (Bottom row) Pixel spectra minus a power law spectrum with spectral index -2.7.  This plot shows the deviations of the added spectral structure from a perfect power law.}
    \label{fig:diffuse-perturbations}
\end{figure*}

To construct the point source catalog for case 3, we merely selected a subset of point sources whose flux at the lowest frequency in our simulations satisfied $S(\nu=100\text{ MHz})\geq15$ Jy.  We then added 5\% Gaussian random flux errors to this bright point sources only catalog to produce the FG catalog used in case 4.  This 5\% flux error was chosen to be of the same order of magnitude as the uncertainty quoted in Table 4 of \citet{hurley17}, i.e.\ $8\pm0.5$\%.  In their work, Table 4 compares the flux values obtained from GLEAM with other overlapping surveys, e.g.\ the VLA Low-frequency Sky Survey Redux \citep[VLSSr;][]{lane14}, the Molonglo Reference Catalogue \citep[MRC;][]{large81, large91}, and the NRAO VLA Sky Survey \citep[NVSS;][]{condon98}.

The FG catalogs for case 5 (a, b, c, and d), were constructed from the true point source catalog (includes all point sources regardless of flux) and a perturbed map of diffuse emission.  We applied both spatial and spectral perturbations to the maps of diffuse emission.  The spectral perturbations were derived from eigenvectors of a frequency-space covariance matrix for a source with a power law spectrum and a tunable decorrelation length $\xi$, of the form \citep{alonso2017}
\begin{equation}
    \mathbf{C}(\nu, \nu') = \left( \frac{\nu\nu'}{\nu_0^2} \right)^{-2.7} \exp\left[ -\frac{\log^2(\nu/\nu')}{2\xi^2} \right]
\end{equation}
For small values of $\xi$, the data are correlated on large spectral scales (or small scales for large values of $\xi$).  Roughly speaking, for our chosen frequency range, the correlation length scales as $\xi_\nu\sim\xi\cdot50$ MHz.  We thus chose four values of $\xi$, 10, 1, 0.1, and 0.01, corresponding to spectral perturbations on $>$ bandwidth scales ($\xi=10$, $\xi_\nu\sim500$ MHz), $\sim$ bandwidth scales ($\xi=1$, $\xi_\nu\sim50$ MHz), $<$ bandwidth scales ($\xi=0.1$, $\xi_\nu\sim5$ MHz), and $\ll$ bandwidth scales ($\xi=0.01$, $\xi_\nu\sim0.5$ MHz).  For each value of $\xi$, we computed the corresponding covariance matrix and derived a set of eigenvectors.  For each eigenvector, we generated a \hpx\ map of spatial perturbations using the \texttt{synfast} function in \texttt{healpy} \citep{healpy}.  \texttt{synfast} generates \hpx\ maps from an input angular power spectrum $C_\ell$.  We chose an input angular power spectrum of $C_\ell = \exp\left( -\ell^2 / \ell_0 \right)$ with $\ell_0=3$ such that the angular power spectrum is heavily suppressed for $\ell\gtrsim10$ ($\lesssim18^\circ$).  This yields a map of spatial perturbations restricted to $\ell\lesssim10$ (spatial scales $\gtrsim18^\circ$). The combined map of spectral and spatial perturbations was then calculated as the sum of each spectral eigenvector times the corresponding map of spatial perturbations. 

\begin{figure}
    \centering
    \includegraphics[width=\linewidth]{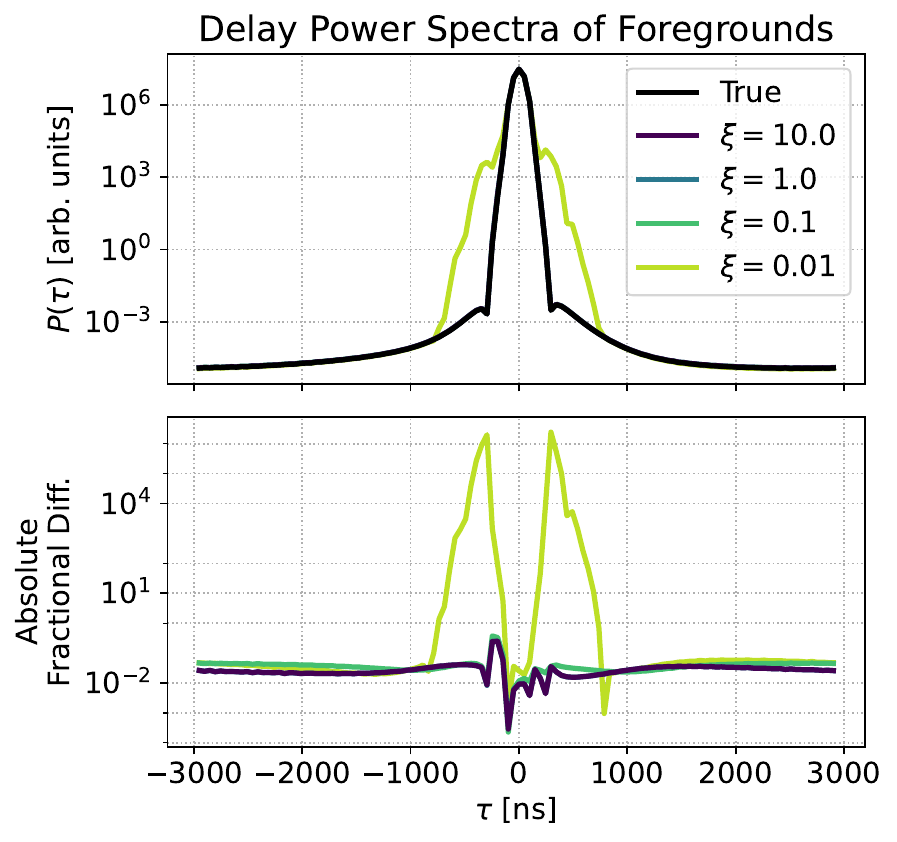}
    \caption{(Top) Delay power spectrum (DPS) of visibilities simulated from the true FGs (black line) and the true point sources + perturbed diffuse catalogs (colored lines, $\xi$ value in legend).  (Bottom) Absolute fractional difference of the DPS of the perturbed and true FG catalog visibilities.  While little difference is visible in the top plot between the true and $\xi\geq0.1$ DPS, we can see deviations at the few percent level across all delays in the absolute fractional difference.  The largest deviation is observed in the $\xi=0.01$ DPS which has significant excess power at $\tau<1000$ ns.}
    \label{fig:dps-fgs-versus-xi}
\end{figure}

The resulting maps of spectral and spatial perturbations were then scaled in amplitude such that the sky-averaged fractional error relative to the unmodified GSM map was 5\%.  This 5\% figure was chosen to be the same order of magnitude as the errors quoted in \citet{gsmref} (see Figure 4 and the supporting text in Section 4.2 in their work).  For the frequencies of interest here (100-120 MHz), \citet{gsmref} claims $\sim5$\% errors in the 2016 GSM catalog which we used here in our visibility simulations.  Examples of these diffuse perturbation maps can be seen in \figref{\ref{fig:diffuse-perturbations}} for $\xi=1$ and 0.01.  The DPS of the visibilities simulated from all of these perturbed diffuse catalogs can be seen in \figref{\ref{fig:dps-fgs-versus-xi}}.  The final maps used in the visibility simulations were calculated as the sum of the unperturbed GSM map and the map of perturbations.

\begin{figure*}
    \centering
    \includegraphics[width=0.9\linewidth]{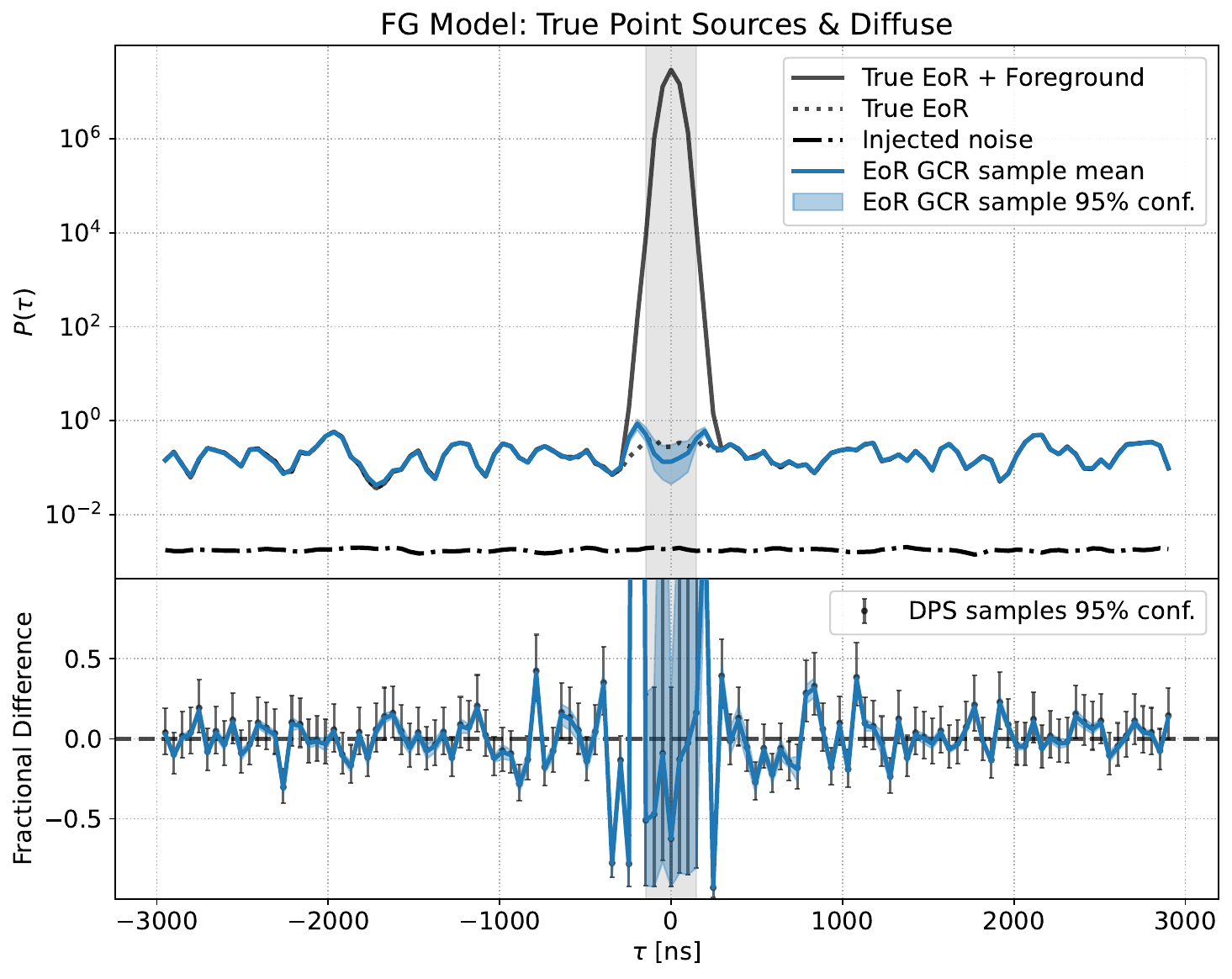}
    \caption{{\it (Top):} Delay power spectrum (DPS) recovery results when using a FG model derived from the true FGs.  The DPS of the true EoR + foregrounds, EoR, and noise are plotted as the solid, dotted, and dashed dotted black lines respectively.  The mean and 95\% confidence interval of the DPS calculated from the GCR samples are plotted as the blue solid line and shaded region, respectively.  The gray shaded region indicates the delay bins which had a uniform prior applied to the delay power spectrum amplitude ($\tau=0$ ns $\pm$ 3 delay bins).  {\it (Bottom):} Fractional difference relative to the true EoR between the GCR samples in blue (GCR / true $-$ 1) and the DPS samples in black (DPS / true $-$ 1).  The blue shaded region and black errorbars represent the 95\% confidence interval of the fractional difference.}
    \label{fig:true-dps-summary}
\end{figure*}

\section{Results}
\label{sec:results}

In this section, we present the results of confronting our fixed simulated data with \hydrapspec\ analyses that assume a variety of different incomplete/perturbed FG models.

\subsection{True FG Model}
\label{sec:results-true-fgs}

\figref{\ref{fig:true-dps-summary}} shows the results from \hydrapspec\ when analyzing data using the ``true'' FG model.  In this case, the FG model is derived from the unmodified FGs in the input data.  The DPS of the input data (mock EoR + FGs + noise), mock EoR, and noise are seen as the solid, dotted, and dashed-dotted black lines in the top subplot, respectively.  The blue line and shaded region show the mean and 95\% confidence interval of the DPS of the GCRs (\evec), respectively.  The shaded gray region indicates the delay bins which had a truncated log-uniform prior applied to the DPS amplitudes.  The bottom plot shows the fractional difference between the outputs of \hydrapspec\ and the DPS of the input mock EoR signal computed as (output / input - 1).  The black data points with errorbars show the fractional difference of the DPS estimates ($P(\tau)_i$) with errorbars corresponding to the 95\% confidence interval.

As previously mentioned, there is a truncated log-uniform prior placed on the central 7 delay bins restricting the range of EoR power spectrum amplitudes and reducing the degeneracy between the EoR and FG models at these low delays.  We can see that the DPS samples diverge from the input EoR DPS outside this symmetric-prior region and where the FGs dominate.  At $\tau\gtrsim300$ ns, where the EoR dominates, we see that the 95\% confidence interval of the DPS samples are consistent with the input EoR DPS.  At low delays, just outside the region where the truncated log-uniform prior is applied, we see that the EoR model diverges from the true EoR signal in the data.  This is visible in \figref{\ref{fig:true-dps-summary}} as the spikes in the DPS of the GCRs (solid blue line) in the top subplot or the large fractional differences in the bottom subplot in the delays bins adjacent to the grey shaded region in the bottom subplot.  This appears to be a FG modelling error of some kind, but the source of this error is still being investigated.  Pinning down the source(s) of this model error has proven difficult due to the degeneracy between the EoR and FG signals at these low delays.  We have run many subsequent tests using mock EoR and FG datasets such as mock FG visibilities derived directly from the spectral templates used in the FG models, \Gmat, and mock EoR and FG signals constructed as sums of delay modes (sinusoids).  The results of these and other tests were unfortunately inconclusive as to where the error lies.  More work is required to address the degeneracy between the EoR and FG signals at low delays.  We wish to point out, however, that in all analyses performed here, the effects of this model error are isolated to low delay modes where FGs dominate the EoR.  It has no affect on our ability to recover the EoR DPS outside of $\tau\gtrsim300$ ns.

\begin{figure*}
    \centering
    \includegraphics[width=\linewidth]{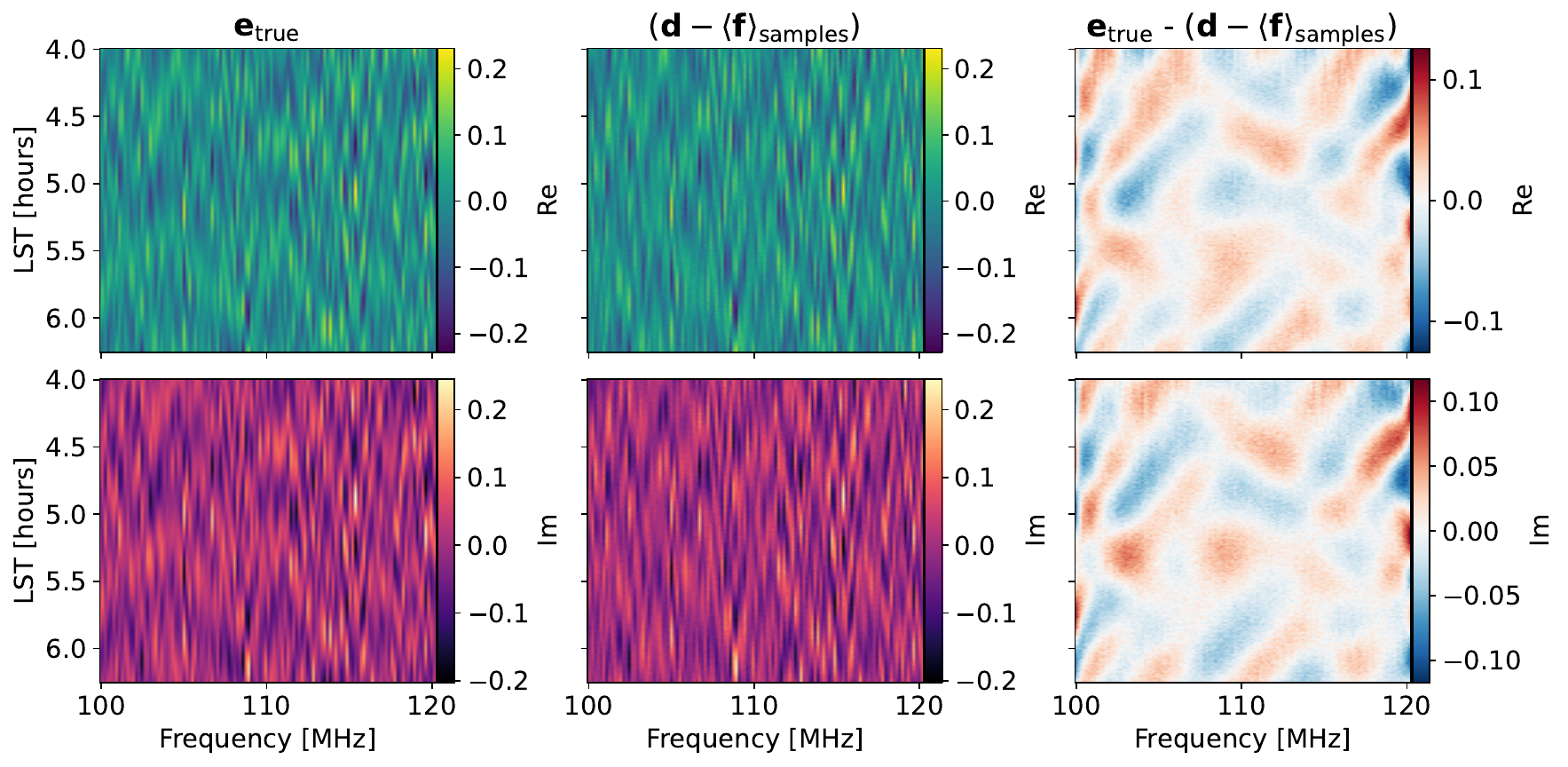}
    \caption{Visibility comparison between the input visibilities and various model components from \hydrapspec.  The top and bottom rows show the real and imaginary components of the visibilities, respectively.  ({\it Left}) Input, mock EoR visibilities ($\evec_{\text{true}}$). ({\it Middle}) Residuals computed as the data ($\dvec = \evec_{\text{true}} + \fvec_{\text{true}} + \nvec$) minus the posterior-weighted FG model ($\left<\fvec\right>_{\text{samples}}$).  ({\it Right}) The difference of the left and middle columns.  If the modelling is working properly, then the residuals $\dvec - \left<\fvec\right>_{\text{samples}}$ should look like the input EoR visibilities (plus fluctuations due to the noise in the data).  The rightmost plot shows us that these residuals do indeed look like the input EoR signal, however, there is evidence of correlations in both frequency and time.  The time correlations are not currently modeled in \hydrapspec but will be included in the next iteration (see \secref{\ref{sec:time-correlations}}).
    }
    \label{fig:vis-residuals}
\end{figure*}

In \figref{\ref{fig:vis-residuals}}, we compare the input and model visibilities.  The top and bottom rows show the real and imaginary components of the visibilities.  The left column shows the input mock EoR visibilities.  The middle column shows the residuals of data minus the posterior-weighted-average FG model.  The data vector in this case is the input visibilities which represent the true sky signals.  If the modelling is working properly, then the difference of data minus FG model should return residuals which look like the input EoR signal (plus noise fluctuations).  The right column differences the left and middle columns.  We see evidence of correlations in the residuals in the right column in both frequency and time, but overall the residuals look quite similar to the mock EoR visibilities.

The success of this test case marks a significant increase in complexity of the data being analyzed as compared to that in \citetalias{kennedy23}.  First, the mock EoR visibilities in \citetalias{kennedy23} were generated as a set of random draws from a Gaussian distribution with mean zero and a known covariance.  These samples were drawn independently for each time to obey the assumption of the analysis.  Here, we instead made mock EoR sky catalogs and {\it simulated} visibilities.  Second, \citetalias{kennedy23} included only point source FG emission in the input data whereas we included both point source and diffuse FG emission.  Third, the data in \citetalias{kennedy23} used a uniform beam which lacks spatial and spectral structure.  Here, we generated a mock EoR sky signal and simulated a set of visibilities using an Airy beam.  This Airy beam has spatial structure that evolves with frequency and introduces additional spectral structure into the simulated visibilities.  For reference, with a dish diameter of 14.6 m, the full-width-at-half-maximum (FWHM) of a Gaussian beam fit to the main lobe of the Airy beam is $\sim12^\circ$ at 100 MHz.  By simulating mock EoR visibilities using this beam, we also introduced time correlations in the data.  For example, the time it takes for a specific right ascension value to pass through the beam used here as the Earth rotates is $\sim 48$ min.  The simulated visibilities thus have non-trivial  temporal correlations on $\sim 48$ min time scales due to the inclusion of the Airy beam (see \secref{\ref{sec:time-correlations}} for more discussion on the topic of time correlations).


Recall, however, that we only used the first 12 principal components in each of our FG models.  Using more principal components can in principle yield a more accurate FG model.  But, as we can see in \figref{\ref{fig:fg-model-pca}}, including more principal components also means including FG templates with increasingly complex spectral structure.  While these higher order principal components may be necessary to fully describe the FGs, by including them in our FG model we risk absorbing EoR power at intermediate to high delays into the FG model.  With this risk in mind, we also performed a test where we reduced the number of FG model basis vectors to minimize the high delay structure in the FG model spectral templates.  When using 9 instead of 12 principal components, the only discernible differences between the two scenarios was again at low delays.  The bounds of the prior at low delays had to be increased by an order of magnitude when using 9 principle components, but the EoR delay power spectrum estimates for $|\tau|\gtrsim300$ ns remained unchanged.  This test further is further evidence that FG model errors are localized to low delays where the FGs dominate the EoR and have no effect on our ability to detect the EoR signal at intermediate to high delays.


\begin{figure*}
    \centering
    \includegraphics[width=\linewidth]{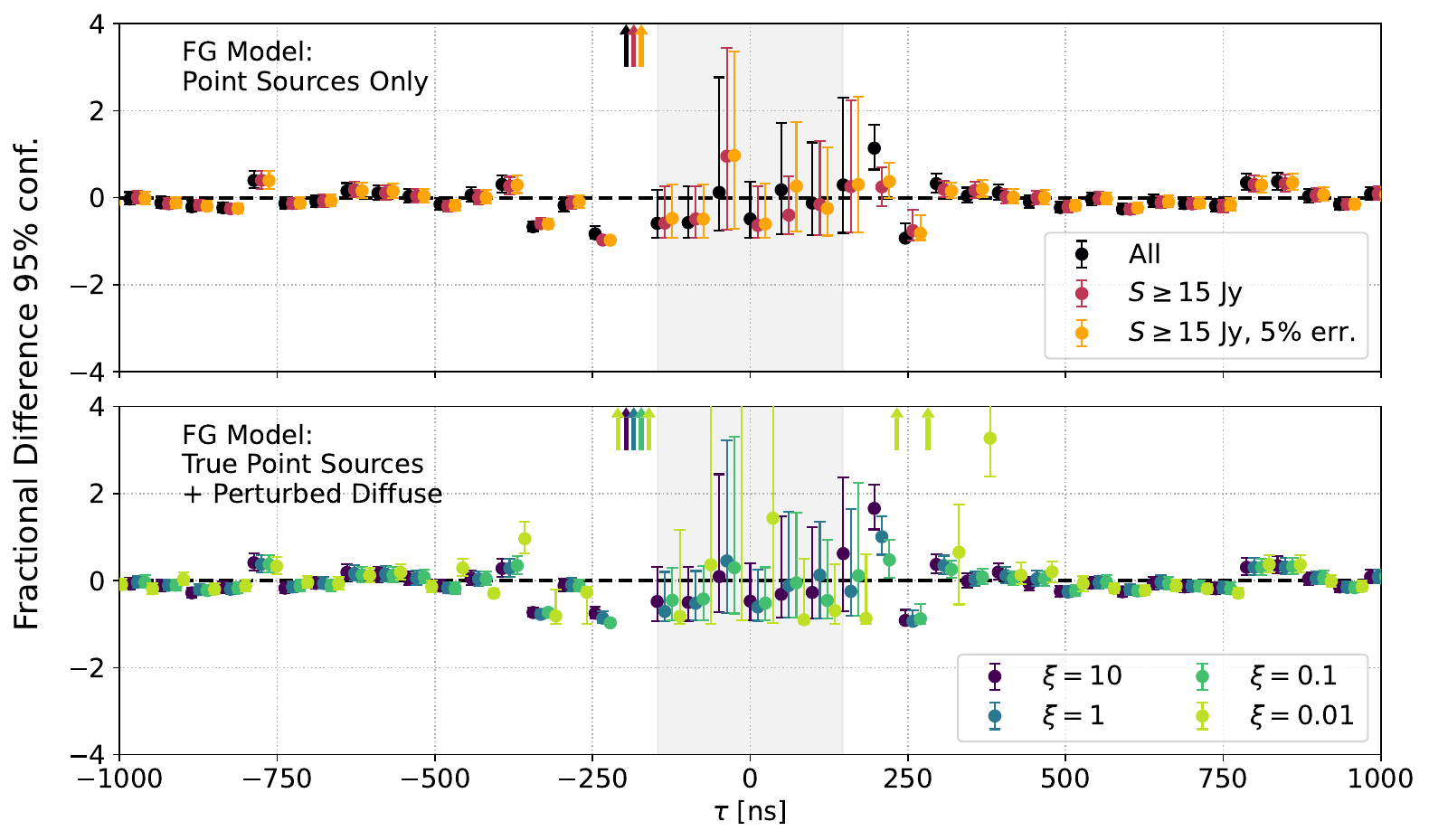}
    \caption{Fractional difference of the recovered delay power spectrum from \texttt{hydra-pspec} relative to the known, input EoR delay power spectrum for each FG model.  The errorbars mark the 95\% confidence interval of each posterior.  The light-gray shaded region marks the delay bins which used a uniform prior on the delay power spectrum amplitude.  The top and bottom subplots show the fractional differences for FG models derived from the point sources only (top) and the true point sources + perturbed diffuse (bottom).  The legend in each subplot details the FG model that was used in each analysis.  Data points whose 95\% confidence intervals lie entirely outside the $y$ range of the plot are indicated with up arrows.  Note that we have zoomed in to $|\tau|\leq1000$ ns to inspect the low delay modes where FG power is concentrated and that the data points have been artificially offset in delay for visual clarity.  Some variation can be seen in the posteriors just outside the prior region but, in a given delay bin, the 95\% confidence regions are all consistent with one another.  Very little variation in the delay power spectrum posteriors is visible for $|\tau|\gtrsim300$ ns where EoR power dominates.  These results demonstrate that no spurious contamination from FG mismodelling is being propagated to higher delays where we aim to detect the EoR.}
    \label{fig:dps-ptsrc-xi}
\end{figure*}

\subsection{Point Source Only FG Models}
\label{sec:results-ps-only-models}

The top subplot of \figref{\ref{fig:dps-ptsrc-xi}} shows the results from \hydrapspec\ when using FG models derived from point source only FG visibilities.  As a reminder, in any test using a modified FG model (Sections \ref{sec:results-ps-only-models} and \ref{sec:results-pert-diff-models}), the input (``true'') data are kept fixed which are simulated from the mock EoR and unmodified FG catalogs (full point source catalog + unperturbed GSM).  These true visibilities represent the sky as seen by the interferometer.  The FG visibilities simulated from the perturbed FG models are used to derive a new set of FG spectral templates, \Gmat, which are used as the FG model to analyze the true visibilities.  We are thus attempting to fit the true FG signals on the sky as seen by the instrument using imperfect FG models.

In \figref{\ref{fig:dps-ptsrc-xi}}, the legend in the top right indicates the FG model used for each set of colored data points.  To reiterate, the data points and errorbars mark the mean and 95\% confidence interval of the fractional difference of the DPS samples (output) and the input EoR DPS computed as (1 $-$ output / input).  Note that we have zoomed in to the delay range $|\tau|\leq1000$ ns.  This is to highlight the small observed differences in the DPS samples as we varied the FG model.  No observable differences in the posteriors of the DPS bins were seen for $|\tau|>1000$ ns.  As in \figref{\ref{fig:true-dps-summary}}, the gray shaded region indicates the delay bins which used a uniform prior on the EoR DPS amplitude.  The only observed differences as we varied the FG model can be seen just outside of this gray shaded region.  Looking back at the top subplot in \figref{\ref{fig:true-dps-summary}}, this region just outside the gray shaded region is where the FGs dominate the EoR signal.  While we can see that the means of the posteriors for these delay bins vary as we change the FG model, the posteriors are all consistent with one another within their 95\% confidence intervals.  For $|\tau|>300$ ns, we see no observable difference in the posteriors for the EoR DPS as a function of the FG model.

The results in \figref{\ref{fig:dps-ptsrc-xi}} indicate that FG models derived from the point source FGs only are sufficient to recover unbiased estimates of the EoR power spectrum for $\tau\gtrsim300$ ns. This is a surprising and promising result given the FG signal in the input data is comprised of both point source and diffuse FG emission and the diffuse emission dominates the point sources in the DPS (see \figref{\ref{fig:dps-sim-components}}).  We can also see that a FG model derived from the brightest point sources only (arbitrarily chosen as those with a flux $\geq15$ Jy) is sufficient to recover the EoR power spectrum for $\tau\gtrsim300$ ns.  This is true even when we incorporate 5\% random flux errors on these brightest point sources.  Note, however, that the impact of an inaccurate/incomplete sky model on calibration at this level could be significant \citep{Barry:2016cpg}.  This is an effect that we do not include in our simulations.  As previously mentioned, more work is required to deal with the degeneracy between the EoR and FG signals at low delay where we do see deviations in EoR power spectrum recovery as a function of the FG model.

\begin{figure*}
    \centering
    \includegraphics[width=\linewidth]{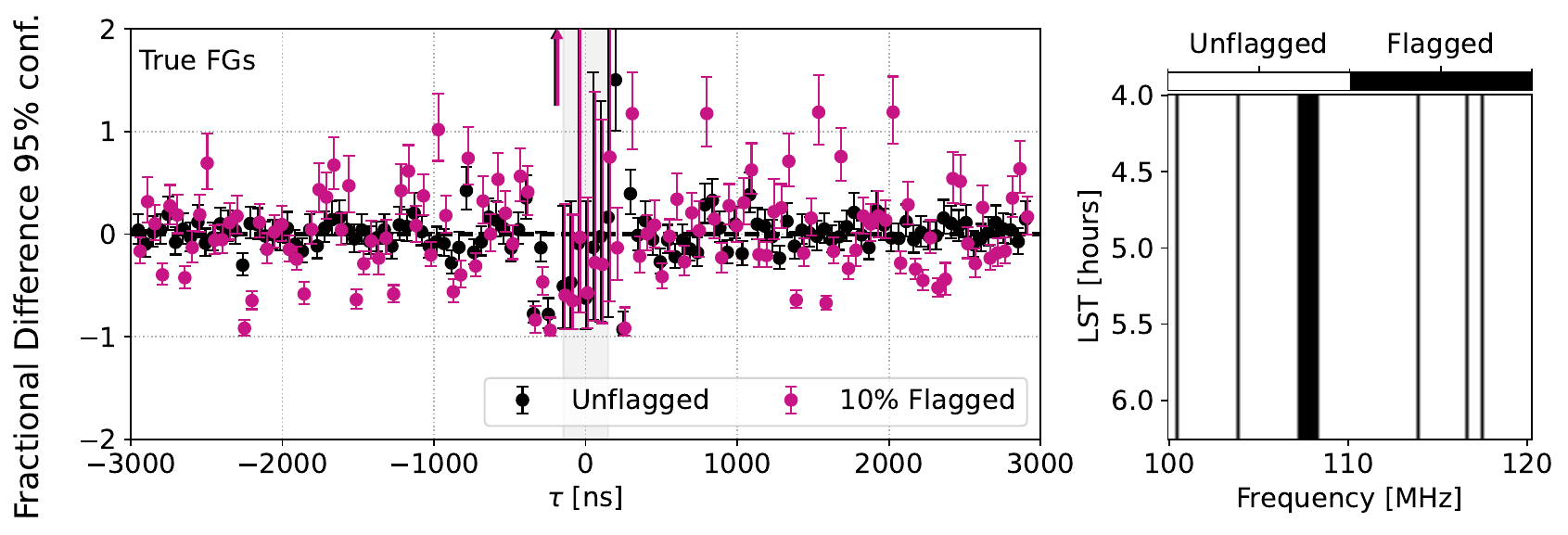}
    \caption{(Left) Fractional difference between the true and recovered delay power spectra with (pink) and without (black) flags when using the true FG model.  As in \figref{\ref{fig:dps-ptsrc-xi}}, data points whose 95\% confidence intervals lie entirely outside the $y$ range of the plot are indicated with up arrows.  (Right) The flagging pattern used for the \hydrapspec\ analyses using flags.  White and black pixels represent data that are unflagged (data is used) and flagged (data is excluded due to contaminants such as RFI), respectively.  These flags result in 10\% of each time sample being flagged.
    }
    \label{fig:dps-with-flags}
\end{figure*}

\subsection{Perturbed Diffuse FG Models}
\label{sec:results-pert-diff-models}

The bottom subplot of \figref{\ref{fig:dps-ptsrc-xi}} shows the results from \hydrapspec\ when using FG models derived from the true point source catalog and modified diffuse catalogs.  The legend in this subplot indicates what value of $\xi$ was used for the FG model in each test case.  Recall that the correlation length $\xi_\nu$ as a function of $\xi$ scales roughly as $\xi_\nu\sim\xi\cdot50$ MHz.  Smaller values of $\xi$ thus correspond to smaller correlations lengths of the spectral perturbations which introduces spectral structure on smaller spectral scales (at higher delay modes).

Similar to the results for the point sources only FG models, we see no substantial differences in the posteriors of the delay bins for $\xi$ values of 10, 1, and 0.1.  We do, however, begin to see significant deviations in the posteriors for $\xi=0.01$.  This case ($\xi=0.01$) represents a worse case scenario where our FG models contain spurious spectral structure on $\ll$ bandwidth scales.  Even in this scenario, however, these FG modelling error effects are isolated to low delays where the contribution of the FGs is non-negligible.  The only observable differences are again only visible in the delay modes adjacent to the gray shade region (the delay bins with a uniform prior applied).  While the means of these posteriors changes a function of FG model, their 95\% confidence intervals are consistent.  These results in conjunction with the results described in the previous section (\ref{sec:results-ps-only-models}) indicate that our approach is robust to FG modelling errors.

\subsection{Data with Flags}
\label{sec:results-flagged-data}

As we stated earlier, an advantage of our approach is the flexibility to use our approach on data with gaps due to flagging (see the paragraph around \eqref{\ref{eq:flag-noise-cov}}).  \figref{\ref{fig:dps-with-flags}} shows the results of a set of tests using flags that are constant in time.  In each time sample, the flagging fraction (percentage of flagged channels within the band) is 10\% with a 7-channel-wide gap near the middle of the bandwidth.  A wide flagging gap near the center of the band represents a worst-case scenario; power spectrum bands are typically chosen to minimize the number of flags, especially wide flagging gaps near the center of the band (see e.g.\ \citet{h1c-validation}).  However, choosing such a challenging scenario gives us a picture of what the most severe effects look like when performing a power spectrum analysis on flagged data. While the results are roughly comparable to those without flags (\figref{\ref{fig:true-dps-summary}}), there is noticeably more scatter in the means of the posteriors of each delay bin relative to the input EoR DPS. This result was also observed in Figure 6 of \citetalias{kennedy23}.

We only show one plot for the test case involving flags in this work to avoid too much duplication.  We specifically tested two FG models when using flags, case 1 (True FGs) and case 5d ($\xi=0.01$).  The results for both cases were identical to the discrepancies seen at low delays in \figref{\ref{fig:dps-ptsrc-xi}}.  We only tested these two scenarios as they represent the extrema of the results we observed.  The true FGs and $\xi=0.01$ FG model DPS results demonstrated the minimum and maximum deviation from the known input EoR DPS.  We do not expect other cases (2$-$4, 5a$-$c) to demonstrate additional adverse effects in comparison to these.

\subsection{Time Correlations}
\label{sec:time-correlations}

As currently implemented, our code assumes that each time sample present in the input data is independent.  The simulated visibilities here, however, are correlated in time due to the beam.  For drift scan observations (like HERA), the antennas point at zenith while the sky drifts overhead as the earth rotates.  As a patch of sky moves through the beam, depending upon the rate at which the data are obtained, the same patch of sky may be present in the main lobe of the beam in many time samples.  For example, for a Gaussian beam with a FWHM of 10$^\circ$, the correlation length in time (or ``beam crossing time'') is 40 minutes as the sky rotates through 1$^\circ$ every 4 minutes.

For the simulations used here, we used an Airy beam corresponding to a dish with a diameter of 14.6 m and a time cadence of 40 s.  The FWHM of a Gaussian fit to the main lobe of this Airy beam is $\sim12^\circ$ at 100 MHz.  This beam has a beam crossing time of $\sim48$ minutes.  At a 40 s observation cadence, $\sim72$ time samples should be correlated due to this beam crossing time.  By assuming each time sample is independent, we are mis-estimating the variance over LST we use to obtain the DPS samples.  In doing so, we are not accounting for the covariance between observations in the data which affects the scale parameter, $\beta$.  Our current focus in the development of \hydrapspec\ is adding a way to model time correlations.  Depending upon the structure of the time covariance matrix of the data, this may be as simple as including a single exponential decay model parameter.  It might also be necessary to include a model of the full time-covariance matrix.  The details of this time-correlation model ultimately require further investigation and is left as future work.

\begin{figure*}
    \centering
    \includegraphics[width=\linewidth]{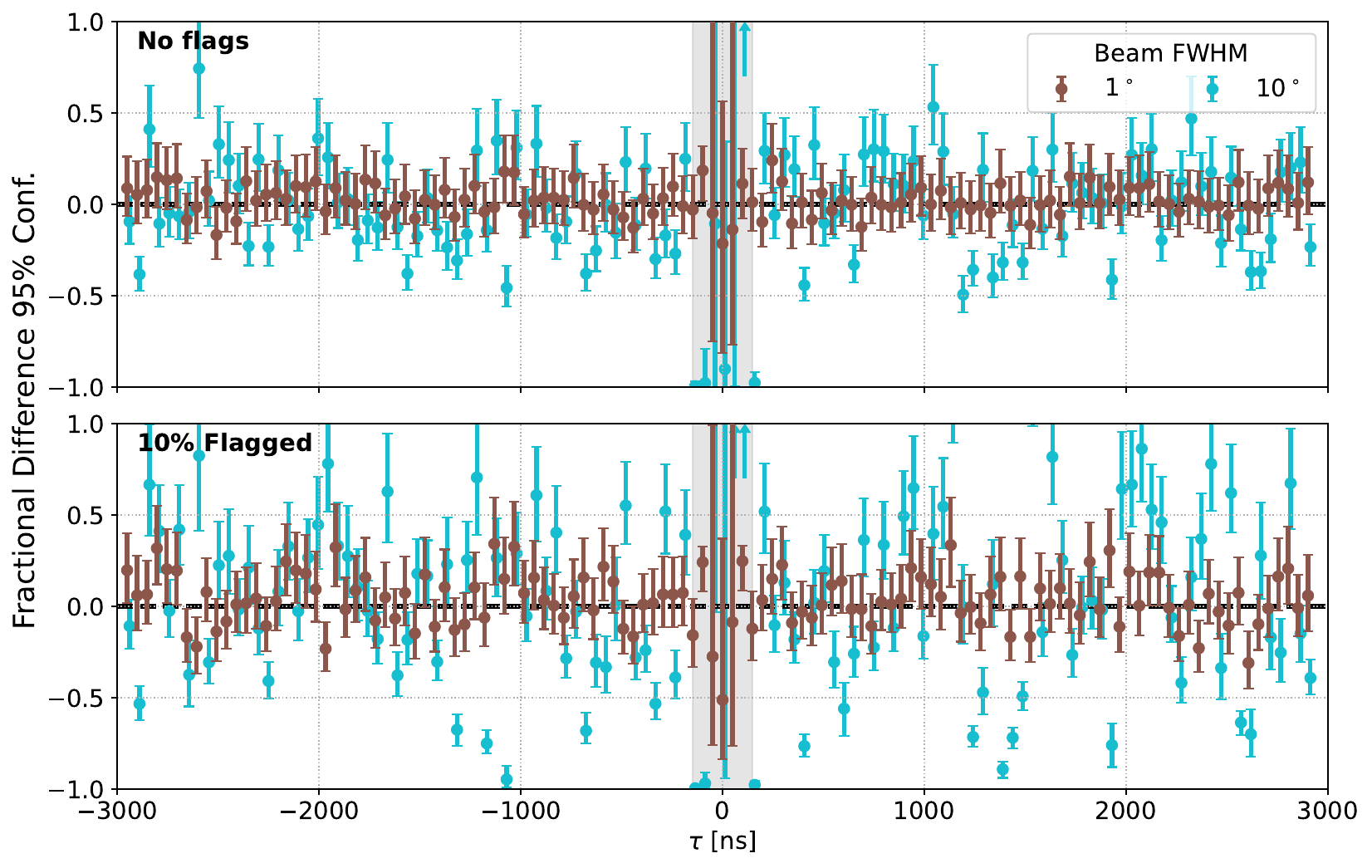}
    \caption{Same as in \figref{\ref{fig:dps-ptsrc-xi}} but for datasets simulated with Gaussian beams using two different FWHM values.  The top and bottom subplots compare analyses running using no flags or flags that are continuous in time and result in 10\% of the band being flagged, respectively.  The width of the beam determines the correlation length in time of the visibilities and a larger beam width results in a longer correlation time (see \secref{\ref{sec:time-correlations}} for more information).  As we narrow the width of the beam, we decrease the correlation time and adjacent LSTs become less correlated with one another.  We see much less scatter in the means of the posteriors for the FWHM=1$^\circ$ beam versus the FWHM=10$^\circ$ beam.  These results demonstrate the importance of modelling time correlations in the visibilities to obtain accurate estimates of the DPS, especially when the data are contain flags with a large flagging gap near the center of the band.}
    \label{fig:dps-vs-fwhm-flags}
\end{figure*}

We did, however, perform a test to elucidate the effects of ignoring these time correlations.  In this test, we simulated two additional datasets that are identical to the simulated data analyzed here except they used a Gaussian beam with FWHM values of 1$^\circ$ and 10$^\circ$.  For the FWHM=1$^\circ$ beam, the beam crossing time is only $\sim$4 minutes and thus the correlation length in the data is 6 time samples (compared to 203 total time samples).  For the FWHM=10$^\circ$, the beam crossing time is $\sim40$ minutes and the correlation length in the data is 60 time samples.  The results of this test can be seen in the top subplot of \figref{\ref{fig:dps-vs-fwhm-flags}}.  The scatter in the means of the posteriors of the delay bins in \figref{\ref{fig:dps-vs-fwhm-flags}} relative to the true EoR DPS is significantly less for the FWHM=1$^\circ$ beam versus the 10$^\circ$ beam. While time correlations are still present in the FWHM=1$^\circ$ data, the correlation length is significantly reduced.  We thus have a larger number of independent samples when compared to the FWHM=10$^\circ$ data.  Our estimates for the shape and scale parameters of the inverse-Gamma distribution are thus closer to their appropriate values. 

While not modelling the time correlations seems to affect the scatter of the posterior means around the true EoR DPS, it does not appear to introduce any bias (see also Figures \ref{fig:true-dps-summary} and \ref{fig:dps-ptsrc-xi}).  The fractional difference plots indicate that the posterior means are scattered symmetrically around the expected EoR DPS.

These time correlations were not an issue in \citetalias{kennedy23} because the EoR ``visibilities'' were drawn as independent samples from a fixed frequency covariance matrix.  There was thus no need to account for any time correlations in the EoR signal in \citetalias{kennedy23}.  Here, however, these time correlations were found to be important because we were starting with a mock EoR sky catalog and simulating visibilities.  It is precisely during observation (simulation in our case) that these time correlations are introduced into the EoR visibilities due to the physical extent of the beam on the sky.

\section{Conclusions}
\label{sec:conclusions}

Spectral (power spectrum) analysis of high-dynamic range data is fraught with difficulties, particularly as real-world effects such as flagging/missing data, inhomogeneous noise, and imperfect knowledge of the dominant foreground component of the signal are encountered. Unless these effects are treated very carefully, spurious couplings of Fourier modes can occur, scattering power outside the foreground-dominated region of Fourier space into the much fainter EoR signal-dominated region. These issues also complicate efforts to estimate the uncertainty on the recovered power spectrum estimate, which is crucial for the correct physical interpretation of the data.

By coupling realistic forward models of all components of the data with Bayesian parameter estimation methods, many of these issues can be addressed, albeit at the expense of increased complexity of the analysis tools. For instance, forward models can extend their predictions into regions of missing data, while recovering the posterior distribution of the model parameters permits uncertainties on all components of the model to be rigorously estimated. The performance of these methods is contingent on the accuracy of the forward model however; model errors (discrepancies between the assumed models and the true behaviour of the underlying physical processes) can lead to increased uncertainties, biases in the recovered parameters, and even spurious effects due to leaving residuals (between the data and best-fit model) with structure beyond what is expected from the noise alone.

\hydrapspec\ is a Bayesian parameter estimation code that models the foregrounds (FGs), the EoR 21~cm signal, and the power spectrum of the latter on a per-baseline basis. In comparison to more traditional optimal quadratic estimator (OQE) methods, it is capable of handling missing data directly, using the method of Gaussian constrained realizations (GCR) to fill-in flagged regions with plausible statistical realisations of the data model. It also permits the covariance of the signal (and therefore the weighting of the data) to be learned as part of the parameter estimation process, rather than needing a fixed a priori model or estimate to be specified. Since the data model also explicitly contains foregrounds and the 21~cm field, \hydrapspec\ can separate them, i.e. it also performs a kind of foreground removal. While in principle this method can use any sensible linear basis to model the foreground and 21~cm signal components, in practice we have use an eigendecomposition of the frequency-frequency covariance matrix of the foreground model visibilities for each baseline to provide the foreground model basis functions. This runs the risk of introducing model errors to the analysis, as the simulations are incomplete (due to imperfect foreground and instrument models), and we have neglected any time dependence of the foreground covariance.

We explored the robustness of \hydrapspec\ to these issues as follows. We first simulated a HERA-like instrument using a sky model consisting of a mock EoR model (based on a simple white noise \hpx\ map at each frequency) and existing FG catalogs (GSM and GLEAM), plus a simplified array layout and an Airy model for the primary beam. We analyzed these same simulated data in \hydrapspec\ using a slew of different assumed FG models. The input data were kept the same in all cases, consisting of the simulations just described, plus thermal noise drawn as white noise visibilities at an amplitude chosen such that the mock EoR signal should be easily detectable (SNR=100) in delay power-spectrum space. For simplicity, we focused only on a single 14.6~m East-West baseline in this work (one of the most numerous, and therefore most sensitive, of HERA's baseline types), although in realistic applications, multiple baselines can be analyzed in parallel if desired. We note that these simulations represent a significant improvement in realism over the ones used in \citetalias{kennedy23}, which made several simplifying assumptions about the primary beam and the temporal structure of the sky model.

The results from \hydrapspec\ in terms of estimated delay power spectrum bandpowers and their associated uncertainties were then compared across the various foreground models assumed in the analysis. Regardless of the FG model we used (point sources only; bright point sources only -- with and without random flux errors; point sources + perturbed diffuse emission), we found that \hydrapspec\ was able to obtain accurate and unbiased estimates of the mock EoR delay power spectrum for $\tau\gtrsim300$ ns (see \figref{\ref{fig:dps-ptsrc-xi}}).  Only in the most extreme scenario (case 5d, $\xi=0.01$) did we observe significant discrepancies in the delay spectrum posteriors at low delays, in the region just outside the delay bins that were dominated by the prior. These results indicate that, at the level of current uncertainties on our FG models, our Bayesian approach to power spectrum estimation is robust to FG modelling errors, although the delay modes close to the edge of the wedge (around the horizon delay) must be treated with care.

We did find, however, that the time correlations in the data, set by the beam-crossing time, play an important role. Our current formulation of the GCR sampling problem does not directly model these time correlations, and essentially assumes that each time sample can be treated as an independent realisation of the data model. This is a conservative assumption as regards recovery of the EoR field and foreground basis function amplitudes, as modelling the correlations would reduce the effective number of degrees of freedom. Conversely, assuming independence of the time samples over-estimates the number of degrees of freedom that go into the delay power spectrum sampling step. Neglecting the time correlations in this way does not impart any overall systematic bias on our delay spectrum estimates, but can result in a mis-estimate of the power spectrum in each delay bin, i.e. a random per-bandpower bias. This is observed in the high-SNR regime that we have studied in this paper, but should be less important in the low-SNR regime where thermal noise dominates. This effect results in an increased scatter in the recovered bandpowers around their true values, which is not captured by the estimated errorbars. This is the case with and without flagging of the data, but becomes more pronounced with a significant flagging fraction.

Incorporating a suitable model for the time correlations in \hydrapspec\ will therefore be an important next step in developing this method, both in order to accurately recover the per-bandpower delay spectrum estimates, and reduce the spurious additional scatter. We conjecture that a simple separable model for the time-and-frequency covariance should improve matters sufficiently, such that $C_{t t^\prime \nu \nu^\prime} \sim C_{\nu \nu^\prime} C_{t t^\prime}$, where $C_{t t^\prime} \sim \exp(-[|t - t^\prime| / T]^\alpha)$ models a simple exponential decay with correlation time $T \sim$~beam-crossing time. The complexity of implementing this kind of model mostly resides in how to efficiently perform the GCR step, as now the per-time GCR samples will be correlated, whereas before each time could be sampled in an embarrassingly parallel fashion. There is also the question of whether an additional Gibbs sampling step should be added to permit sampling of the correlation time $T$ and index $\alpha$, or whether these could be fixed a priori.


\section*{Acknowledgments}

This result is part of a project that has received funding from the European Research Council (ERC) under the European Union's Horizon 2020 research and innovation programme (Grant agreement No. 948764; PB). PB acknowledges support from STFC Grant ST/T000341/1. We acknowledge use of the following software: {\tt corner} \citep{corner}, {\tt matplotlib} \citep{matplotlib}, {\tt numpy} \citep{numpy}, and {\tt scipy} \citep{2020SciPy-NMeth}.  This work used the DiRAC@Durham facility managed by the Institute for Computational Cosmology on behalf of the STFC DiRAC HPC Facility (www.dirac.ac.uk). The equipment was funded by BEIS capital funding via STFC capital grants ST/P002293/1, ST/R002371/1 and ST/S002502/1, Durham University and STFC operations grant ST/R000832/1. DiRAC is part of the National e-Infrastructure.

\section*{Data Availability}

The Python code used to produce the results in this paper is available from \url{https://github.com/HydraRadio/hydra-pspec}. Simulated data are available on request, but can be regenerated using publicly-available tools, e.g. {\tt hera\_sim} (\url{https://github.com/HERA-Team/hera_sim}) and {\tt pyuvsim} (\url{https://github.com/RadioAstronomySoftwareGroup/pyuvsim}).

\balance



\bibliographystyle{mnras}
\bibliography{pspeccov}




\bsp	
\label{lastpage}
\end{document}